\documentclass{article}

\usepackage{amsmath, amssymb, graphicx}
\usepackage[mathscr]{eucal}

\newcommand{\ba}{\begin{array}}
\newcommand{\ea}{\end{array}}

\newcommand{\p}{\partial}
\newcommand{\ud}{\mathrm{d}}
\newcommand{\pathD}{\!\mathscr{D}}
\newcommand{\Det}{\text{Det}\,}

\newcommand{\x}{{\bf x}}
\newcommand{\y}{{\bf y}}

\newcommand{\pp}{{\bf p}}
\newcommand{\kk}{{\bf k}}

\def\Journal#1#2#3#4{{#1} {\bf #2} (#3) #4}

\title{Radial evolution in anti-de Sitter spacetime}
\author{A. Ilderton \\ Centre for Particle Theory, University of Durham, \\ Durham DH1 3LE, UK. \\ \\ a.b.ilderton@dur.ac.uk}
\date{}

\begin{document}
\maketitle
\abstract{Properties of Green's functions may be derived in either first or second quantisation. We  illustrate this with a factorisation property for propagators in arbitrary spacetimes, and apply it to scalar fields in AdS space.}
\subsection*{General results}
Consider $D+1$ dimensional spacetime with metric $g_{\mu\nu}$ and let $\nabla^2$ be the Laplacian,
\begin{equation}
  \nabla^2 = -\frac{1}{\sqrt{g}}\p_a\, \sqrt{g}g^{ab}\p_b.
\end{equation}
Suppose we have a region $V$ of spacetime bounded by a surface $\p V$. Let $\phi$ be a solution to Poisson's equation in $V$,
\begin{equation*}
  \nabla^2 \phi(x) = \rho(x) \quad\forall x\in V
\end{equation*}
for some function $\rho$. Let $G$ be a Green's function for Poisson's equation obeying
\begin{equation*}
  \nabla^2 G(q,z) = \delta(q-z)/\sqrt{g(z)},
\end{equation*}
which holds for $z$, $q$ both within and in some neighbourhood of $V$ and the boundary, then we have
\begin{equation}
  \phi(x)\nabla^2G(x,z) - G(x,z)\nabla^2\phi(x) = \phi(x)\delta(x-z)/\sqrt{g(z)} - G(x,z)\rho(x).
\end{equation}
Integrating both sides over $V$ and applying Green's theorem gives
\begin{equation*}\begin{split}
  \int\limits_{\p V}\!\ud y\,\, G(y,z)&\frac{\p}{\p n}\phi(y)-\phi(y)\frac{\p}{\p n} G(y,z) \\
  &= \int\limits_V\!\ud x\sqrt{g(x)}\,\,\phi(x)\delta(x-z) - \int\limits_V\!\ud x\sqrt{g(x)}\,\,G(x,z)\rho(x)
\end{split}\end{equation*}
where $\ud y$ is a measure on $\p V$ and $\p/\p n$ is the normal derivative on the boundary,
\begin{equation}
  \frac{\p}{\p n} = n^\mu\frac{\p}{\p x^\mu}\bigg|_{\p V}.
\end{equation}
Now choose $z\in V$ and let $\phi(x) = G(q,x)$ for some $q\not\in V$, so that $\rho(x)=\delta(x-q)$ which vanishes in our integration range. Substituting in the above we find
\begin{equation}\label{factor}
  \int\limits_{\p V}\!\ud y\,\, G(q,y)\overleftrightarrow{\frac{\p}{\p n}}G(y,z)= G(q,z).
\end{equation}
where $\leftrightarrow = \leftarrow - \rightarrow$. A simple extension shows that the derivation above also holds when $\phi$ and $G$ are a solution and Green's function of the Helmholtz equation $\nabla^2 + m^2 =\rho$, rather than Poisson's equation. As applied to a scalar field theory this implies that the field propagator may be factorised. This result, in certain situations, also applies to the propagator of bosonic string theory, and has been applied to the construction of the vacuum wave functional \cite{me2} and investigating T-duality \cite{us1} \cite{us2}.

It is interesting to note that this property can be derived from considering first quantised particles. The off-shell free space propagator from the point $x_i$ to the point $x_f$ may be written as a sum over all paths from $x_i$ to $x_f$ equipped with all possible worldline metrics $e$ \cite{Brink} \cite{PolyakovBook},
\begin{equation}\label{prop-path}
  G(x_f; x_i) = \int\pathD (x,\,e)\,\, e^{-\int_0^1 \ud\xi\sqrt{e}\,\, ({\dot x\cdot\dot x}/(2e)+m^2/2)}\bigg|_{x(0)=x_i}^{x(1)=x_f} \\
  \end{equation}
The path is parameterised by $\xi$ and $x(0)$ is the point $x_i$, $x(1)$ the point $x_f$. The intrinsic metric $e$ transforms as a (0,2) tensor,
\begin{equation*}
  e(\xi) = \bigg(\frac{\ud\xi'}{\ud\xi}\bigg)^2e(\xi'),
\end{equation*}
so the propagator is manifestly reparametrisation invariant\footnote{The over counting of equivalent paths is removed by dividing by the volume of the space of reparametrisations, we will not repeat the explicit evaluation of the integrals as this is well documented, see for example \cite{PolyakovBook}.}. This is the particle analogue of the Polyakov integral in string theory. 

Now, with reference to the preceding second quantised discussion, consider a particle path from $z\in V$ to $q\not\in V$. The derivation of the property (\ref{factor}) begins with the simple observation that at some point the path must cross $\p V$ at some (intrinsic, not physical) time. This implies the sum over all paths defining the propagator can factorised so that formally
\begin{equation*}
\sum_{\textrm{paths }z\rightarrow q} e^{-\textrm{length }z\rightarrow q}= \sum\limits_y  \bigg( \sum_{\textrm{paths }z\rightarrow y}
e^{-\textrm{length }z\rightarrow y)}\bigg) \bigg( \sum_{\textrm{paths }y\rightarrow q}
e^{-\textrm{length }y\rightarrow q)}\bigg) 
\end{equation*}
where $y\in\p V$. To make this factorisation explicit, insert into (\ref{prop-path}) a resolution of the identity,
\begin{equation}\label{prop-trick}
	1=\int\!\ud\xi'\sqrt{e(\xi')}\,\, J(x)\,\delta\big(F(x(\xi'))\big)
\end{equation}
where $F(x)=0$ defines the surface $\p V$. This delta function has support for any path $x^\mu(\xi)$ as we have described. $e$ is here an arbitrary worldline metric which will soon drop out. Taylor expanding $F(x)$ the Jacobian $J$ which makes the insertion unity is easily found to be
\begin{equation*}
  J=\frac{\nabla_\mu S\p_\xi x^\mu(\xi)}{\sqrt{e(\xi)}}\bigg|_{S(x(\xi))=0}.
\end{equation*}
Note that this Jacobian is reparametrisation invariant. Taking the integral over $\xi'$ outside and distinguishing between worldline times earlier and later than $\xi'$, the $x^\mu$ integrals in the propagator can be written
\begin{equation*}\begin{split}
	\int\!\ud\xi'\sqrt{e(\xi')}\,\left[\prod_{\xi<\xi'}\int\!\ud x^\mu(\xi)\right]\int\!\ud x^\mu(\xi')\,&J\,\delta\big(F(x)\big)\left[ \prod_{\xi>\xi'} \int\!\ud x^\mu(\xi)\right] \\ 
	&\exp\left(-\sum\limits_{\xi<\xi'} S[x(\xi)] - \sum\limits_{\xi>\xi'} S[x(\xi)]\right).
\end{split}\end{equation*}
Integrating over the delta function gives
\begin{equation*}
  \int\limits_{\p V}\!\ud y \frac{J}{|\nabla F|} = \int\limits_{\p V}\ud y\,\, n_\mu\frac{\partial_\xi x^\mu}{\sqrt{e}}.
\end{equation*}
We can write $\partial_\xi x(\xi)$ as a two sided derivative
\begin{equation*}
  \p_\xi x^\mu(\xi')= \lim_{h\rightarrow 0} \frac{x^\mu(\xi'+h)- x^\mu(\xi'-h)}{2h}
\end{equation*}
which splits the path integration into a pair of terms of two products, one of which has an insertion. Everything is invariant under reparametrisations of the worldline and so have no explicit $\xi'$ dependence, the integral over which gives a finite volume, leaving
\begin{equation}
\begin{split}
	G_0=\frac{1}{2}\int\limits_{\p V}\!\ud y\int\pathD (x,e)\,\, &\frac{n.\dot{x}(\xi_\text{final})}{\sqrt{e(\xi_\text{final})}}e^{-S[x]} \int\pathD (x,e)\,\,e^{-S[x]} \\
	+ \frac{1}{2}\int\limits_{\p V}\!\ud y\int\pathD (x,e)\,\, e^{-S[x]}& \int\pathD (x,e)\,\,\frac{n.\dot{x}(\xi_\text{initial})}{\sqrt{e(\xi_\text{initial})}}e^{-S[x]}
\end{split}
\end{equation}
It can be seen from an integration by parts in the action that the insertions of the Jacobian can be taken outside the integrals as derivatives with respect to boundary data,
\begin{equation*}
  \frac{\p}{\p x^\mu(\xi_\text{final})} = -\frac{1}{2}\frac{\dot{x}^\mu(\xi_\text{final})}{\sqrt{e(\xi_\text{final})}}
\end{equation*}
or with an overall plus sign for the initial time, recovering the result (\ref{factor}).
\subsection*{Example - AdS space}
Let us give an explicit example which may be of interest. Since its proposal \cite{Maldec} there has been huge interest in studying the correspondence between string theory in anti-de Sitter space and its holographic dual conformal field theory on the boundary (\cite{review} for a comprehensive review). A variety of new approaches to studying the correspondence have recently been employed (see for example \cite{Berkovits} and \cite{Bena}...\cite{Hatsu}). Berkovits' approach \cite{Berkovits2} gives a quantisable action for the superstring in AdS space times a compact manifold, with Ramond Ramond flux, but the complexity of the action makes hard work of calculating scattering amplitudes \cite{Berkovits3}...\cite{Triv}.

Given the difficulties of string calculations in anti-de Sitter space, it is worthwhile to pursue unconventional approaches. We believe the factorisation property discussed plays a crucial role in defining the radial dependence of scalar fields in AdS, much as we found for time evolution in flat space \cite{us1} \cite{us2}. Translation invariance simplified the flat space calculations so we do not expect our Feynman diagram arguments to be so simple in AdS, but we can derive the essential property.

We will represent Euclideanised anti-de Sitter spacetime as the upper half space $x^0>0$ with metric
\begin{equation}\begin{split}
  \ud s^2 &= \frac{1}{{x^0}^2}\big(\ud x^0\ud x^0+\ud\x^i\ud\x^i\big)=: A_{ab}\ud x^a \ud x^b, \\
  \quad A(x^0)&:=\Det A_{ab} = \bigg(\frac{1}{x^0}\bigg)^{D+1}
\end{split}\end{equation}
where $i=1\ldots D$ and $x^0$ is the `radial' direction. For a scalar field of mass $m$ define $\nu=\sqrt{ D^2/4 +m^2}$, then the scalar field propagator which vanishes on the boundary of the spacetime (compactified $\mathbb{R}^D$) is, labelling the final and initial values of $x^0$ as $r_f$ and $r_i$ respectively \cite{Burgess} \cite{Muck},
\begin{equation}\label{AdS-prop}
  G_\text{AdS}(r_f,\x_f;r_i,\x_i) = -\int\!\frac{\ud^D\kk}{(2\pi)^D}\, (r_f r_i)^{D/2}e^{-i\kk.(\x_f-\x_i)}K_\nu(|\kk|r_f)I_\nu(|\kk|r_i)
\end{equation}
when $r_f>r_i$ and the Bessel functions $I_\nu$ and $K_\nu$ \cite{Grad} are exchanged otherwise. Now let $V$ be the space $x^0<r$, let $z=(r_i,\x_i)$ and let $q(r_f,\x_f)$ with $r_f >r > r_i$. Our general result implies that the propagator factorises as
\begin{equation}\label{AdS-fact}\begin{split}
  G_\text{AdS}(r_f,\x_f;r_i,\x_i)= \int\!\ud^D\y A(r)\,\,&\bigg(r^2\frac{\partial}{\partial r}G_\text{AdS}(r_f,\x_f;r,\y)\bigg)G_\text{AdS}(r,\y;r_i,\x_i) \\
  &- G_\text{AdS}(r_f,\x_f;r,\y)\bigg(r^2\frac{\partial}{\partial r}G_\text{AdS}(r,\y;r_i,\x_i)\bigg).
\end{split}\end{equation}
To prove this directly is a simple matter. Inserting the explicit representation of the propagator (\ref{AdS-prop}) into the right hand side of (\ref{AdS-fact}) we find
\begin{equation}\begin{split}
  -\int\!\frac{\ud^D(\kk,\pp,\y)}{(2\pi)^{2D}} &(r_f r_i)^{D/2}\,\,e^{-i\kk.(\x_f-\y)}e^{-i\pp.(\y-\x_i)}K_\nu(|\kk|r_f)I_\nu(|\pp|r_i) \\
  &\times r^{1-D/2}\bigg\{ K_\nu(|\pp|r)\frac{\partial}{\partial r} r^{D/2}I_\nu(|\kk|r) - I_\nu(|\kk|r)\frac{\partial}{\partial r} r^{D/2}K_\nu(|\pp|r)\bigg\}.
\end{split}\end{equation}
The integral over $\y$ gives a momentum conserving delta function which allows us to do the integral over $\pp$, say. For brevity write $s\equiv|\kk|r$ and the result of these integrations is
\begin{equation}\begin{split}\label{big}
   -\int\!\frac{\ud^D\kk}{(2\pi)^D}&(r_f r_i)^{D/2}\,\,e^{-i\kk.(\x_f-\x_i)} K_\nu(|\kk|r_f)I_\nu(|\kk|r_i) \\
  &\times s^{1-D/2}\bigg\{ K_\nu(s)\frac{\partial}{\partial s} s^{D/2}I_\nu(s) - I_\nu(s)\frac{\partial}{\partial s} s^{D/2}K_\nu(s)\bigg\} \\
  = -\int\!\frac{\ud^D\kk}{(2\pi)^D}&(r_f r_i)^{D/2}\,\,e^{-i\kk.(\x_f-\x_i)} K_\nu(|\kk|r_f)I_\nu(|\kk|r_i) \\
  &\times s^{1-\nu}\bigg\{ K_\nu(s)\frac{\partial}{\partial s} s^\nu I_\nu(s) - I_\nu(s)\frac{\partial}{\partial s} s^\nu K_\nu(s)\bigg\}
\end{split}\end{equation}
plus two terms coming from the derivatives of $s^{D/2}$ which cancel. We require the final line of the above to be unity in order to recover $G_\text{AdS}$. Applying the Bessel function properties \cite{Grad}
\begin{equation}
  \frac{\partial}{\partial s}s^\nu I_{\nu}(s) = s^\nu I_{\nu-1}(s),\qquad \frac{\partial}{\partial s}s^\nu K_{\nu}(s) = -s^\nu K_{\nu-1}(s), 
\end{equation}
the final line of (\ref{big}) becomes
\begin{equation}
  sI_{\nu-1}(s)K_\nu(s)+sI_\nu(s)K_{\nu-1}(s) = 1,
\end{equation}
a standard Bessel function identity for unity, leaving
\begin{equation}
  -\int\!\frac{\ud^D\kk}{(2\pi)^D}(r_f r_i)^{D/2}\,\,e^{-i\kk.(\x_f-\x_i)} K_\nu(|\kk|r_f)I_\nu(|\kk|r_i) \equiv G_\text{AdS}(r_f,\x_f;r_i,\x_i).
\end{equation}
We have proven (\ref{AdS-fact}). For $r_f<r<r_i$ the right hand side of (\ref{AdS-fact}) picks up a minus sign. These results also hold in the limit in which $r_f=r$ or $r=r_i$.

In the AdS/CFT correspondence translations in the bulk radial direction correspond to conformal transformations in the boundary field theory. Since the correspondence holds between scalar field theory in $AdS_{D+1}$ and conformal field theory on $\mathbb{R}^D$ \cite{WittenAdS}, there is an opportunity to study how our results relate to a holographic description in terms of scaling of operators in the boundary conformal field theory. Toward this end we note that our results hold for the propagator
\begin{equation}\begin{split}
  G_\epsilon(r_f,\x_f;r_i,\x_i) &:= G_\text{AdS}(r_f,\x_f;r_i,\x_i) \\
   &\hspace{5pt}+ \int\!\frac{\ud^D\kk}{(2\pi)^D}(r_f r_i)^{D/2}e^{-i\kk.(\x_f-\x_i)}K_\nu(|\kk|r_f)K_\nu(|\kk|r_i)\frac{I_\nu(|\kk|\epsilon)}{K_\nu(|\kk|\epsilon)}
\end{split}\end{equation}
considered in \cite{Muck} which vanishes not on the spacetime boundary but on the near-boundary surface $x^0=\epsilon$ and is used to regulate boundary divergences when investigating the conformal field theory.


\begin{thebibliography}{99}
\bibitem{me2} A.~Ilderton, ``The vacuum state functional of interacting string field theory", hep-th/0506173

\bibitem{us1} A.~Ilderton and P.~Mansfield, ``Time evolution in string field theory and T-duality" Phys.Lett.{\bf B607}(2005) p294 [hep-th/0410267]

\bibitem{us2} A.~Ilderton and P.~Mansfield, ``Timelike T-duality in the string field Schr\"odinger functional" [arxiv:hep-th/0411166]

\bibitem{Brink} L. Brink, P. Di Vecchia, P. Howe ``A locally supersymmetric and reparametrisation invariant action for the spinning string", \Journal{Phys.Lett.}{B65}{1976}{471}

\bibitem{PolyakovBook}
A.M. Polyakov, ``Gauge fields and strings'', Harwood Academic Publishers (1987) ISBN 3-7186-0492-6

\bibitem{Maldec}J.M. Maldacena, ``The large $N$ limit of superconformal field theories and supergravity'', \Journal{Int.J.Theor.Phys.}{38}{1999}{1113} hep-th/9711200

\bibitem{review} O. Aharony et al, ``Large N field theories, string theory and gravity'', hep-th/ \Journal{Phys.Rept.}{323}{2000}{183}, hep-th/9905111

\bibitem{Berkovits} N.Berkovits, ``ICTP lectures on covariant quantization of the superstring'', published in `Trieste 2002, Superstrings and related matters', hep-th/0209059 

\bibitem{Bena} I. Bena, J. Polchinski, R. Roiban, ``Hidden symmetries of the AdS$_5\times\mathbb{S}^5$ superstring'', \Journal{Phys.Rev.}{D69}{2004}{046002}, hep-th/0305116

\bibitem{Dolan1} L.~Dolan, C.~R.~Nappi, E.~Witten, ``Yangian symmetry in D = 4 superconformal Yang-Mills theory'', hep-th/0401243.

\bibitem{Hatsu} M.~Hatsuda and K.~Yoshida, ``Classical integrability and super Yangian of superstring on AdS$_5\times\mathbb{S}^5$'', hep-th/0407044.

\bibitem{Berkovits2} N. Berkovits ``Super-Poincar\'e covariant quantization of the superstring'', \Journal{JHEP}{0004}{2000}{018}, hep-th/0001035

\bibitem{Berkovits3} N. Berkovits, ``Superstring vertex operators in an AdS$_5\times\mathbb{S}^5$ background'' \Journal{Nucl.Phys.}{B596}{2001}{185}, hep-th/0009168;

\bibitem{Dolan} L. Dolan, E. Witten, ``Vertex operators for AdS3 background with Ramond Ramond flux'', \Journal{JHEP}{9911}{1999}{003}, hep-th/9910205

\bibitem{Bob} K. Bobkov, L. Dolan ``Three-graviton amplitude in Berkovits-Vafa-Witten variables'', \Journal{Phys.Lett.}{B537}{2002}{155}, hep-th/0201027
 
\bibitem{Triv} G. Trivedi, ``Correlation functions in Berkovits' pure spinor formulation'' \Journal{Mod.Phys.Lett.}{A17}{2002}{2239}, hep-th/0205217

\bibitem{Burgess} C.P.Burgess and C.A.L\'utken, ``Propagators and effective potentials in anti-de Sitter space'', \Journal{Phys.Lett.}{B135}{1985}{137}

\bibitem{Muck} W. M\'uck and K.S.Viswanathan, ``Conformal field theory correlators from classical scalar field theory on anti-de Sitter space'', \Journal{Phys.Rev.}{D58}{1998}{041901}

\bibitem{Grad} I.S.Gradshteyn and I.M.Ryzhik, ``Table of integrals, series and products'' ISBN 0-12-294760-6

\bibitem{WittenAdS} E.~Witten, ``Anti-de Sitter space and holography'', Adv.\ Theor.\ Math.\ Phys.\  {\bf 2}, 253 (1998) [arXiv:hep-th/9802150]
\end{thebibliography}
\end{document}